# Phenomenological approach of the thermodynamic properties of CDW (SDW) systems

**M. Saint-Paul*** **and P. Monceau**

Univ. Grenoble Alpes, CNRS, Grenoble INP[a], Institut Néel, 38042 Grenoble, France

## Abstract

The microscopic description of the CDW phase transition is still debated and remains controversial. The question is how to extend the Peierls picture to real systems in higher dimensions. A general tendency is found in the thermodynamic properties such as the specific heat jump $\Delta C_p$ and the decrease of the longitudinal elastic stiffness constant $\Delta C_{11}/C_{11}$ at the CDW phase transition in several materials, such as quasi-one dimensional ($K_{0.3}MoO_3$), transition metal dichalcogenide compounds (2H-$NbSe_2$), rare earth tritellurides ($TbTe_3$, $ErTe_3$, $HoTe_3$) and intermetallic compound ($Lu_5Ir_4Si_{10}$). $\Delta C_p$ and $\Delta C_{11}/C_{11}$ increase as the temperature of the phase transition $T_{CDW}$ and $T_{CDW}^2$ respectively. The same tendency is found at the spin density phase transition in chromium and $CuGeO_3$.Thermodynamic properties of almost all CDW systems, although it has been recognized to exhibit large fluctuations, follow the classical mean field BCS type behavior.

[a]*Institute of Engineering Univ. Grenoble Alpes*

*michel.saint-paul@neel.cnrs.fr

## 1. Introduction

The concept of a charge density wave which is induced by Fermi surface nesting originated from the Peierls idea of electronic instabilities in purely one dimensional metal is now often applied to charge ordering in real low dimensional materials. The microscopic origin of the CDW phase transition in this material is still debated and remains controversial [1-27]. In general a real material does not go through a pure metal insulator transition at $T_{CDW}$ and implies a partial gaping of the Fermi surface. The claim that this transition is a CDW by the community needs to be completely rethought. Transition metal dichalcogenides (TMD) have been the center of CDW research since several decades but the results are still confusing [5-10]. Transitions in the TMD are not CDWs in the sense of a Peierls picture [8]; even in these layered materials they are in some cases 3D in nature. 1T-$TiSe_2$ undergoes a second order phase transition to a commensurate lattice distortion 2x2x2 [17, 18]. Some authors described $TiSe_2$ as a CDW. In fact this transition is a commensurate structural transition which induces periodic charge modulation. The question is how to extend the Peierls picture to real systems in higher dimension.

The elastic constants are thermodynamic derivatives. They are important together with the specific heat and thermal expansion for the equation of state of a material [36-37]. Following our survey of the thermodynamic properties of the charge density wave systems [34] we reexamine the Landau phenomenological approach [34]. As we demonstrate below it appears that both CDW and SDW phase transitions can be described with similar parameters in the mean field approximation [1]. The amplitude of the specific heat at the CDW phase transition increases with increasing $T_{CDW}$ and a tendency to a lineal temperature dependence is verified. In the same manner the amplitude of the discontinuity of the longitudinal elastic constants increases with increasing $T_{CDW}$, a quadratic dependence is verified. Departures from the mean field behavior of the thermodynamic properties are generally attributed to fluctuations which belong to the 3D XY criticality class [1].

## 2 Thermodynamic properties

## *2.2 Landau theory of a second order phase transition*

The physics of quasi one dimensional solids which undergoes a Peierls transition was discussed in terms of a Landau phenomenological approach by Mc Millan for the charge density waves in transition metal dichalcogenides with charge density as an order parameter [34]. The quantities of interest are the temperature at which the transition occurs $T_{CDW}$ and the order parameter. In the case of a spin density wave transition, the thermodynamic quantities near $T_{SDW}$ can be described in terms of the Landau theory which follows the description of the charge density wave [1]. As for a charge density wave the development of the SDW ground state opens up a gap at the Fermi level leading to a metal insulator transition [1].
The free energy F of the system is expanded in power of the order parameter Q in the mean field approximation at a second order phase transition:

$$F(\Delta,T) = F(0,T) + \frac{1}{2}a(T - T_{CDW})Q^2 + \frac{1}{4}cQ^4 \qquad (1)$$

where the coefficients a and c are to be estimated. The order parameter is associated to the gap energy. The order parameter that minimizes the free energy in eq. 1 is $Q = \sqrt{\frac{a}{c}}[T_C - T]^{0.5}$.

which gives $Q(0) = \sqrt{\frac{aT_{CDW}}{c}}$ at T=0. The specific heat jump $\Delta C_P$ at a second order phase transition deduced from eq. 1 is given by

$$\frac{\Delta C_P}{V_m} = \frac{a^2 T_{CDW}}{c} \qquad (2)$$

$V_m$ is the molar volume. An example of the specific jump $\Delta C_P$ at the CDW phase transition in $TbTe_3$ is given in Fig. 1(a). The coefficients a and c were calculated by Allender et al. [35] in one dimensional chain model using a weak coupling mean field description and they are found to be:

$$a = \frac{N(E_F)}{T_{CDW}} \quad \text{and} \quad c = \frac{0.1N(E_F)}{k_B^2 T_{CDW}^2} \qquad (3)$$

These values are nearly identical with the BCS case where $N(E_F)$ is the density of state per unit energy interval at the Fermi level. It is interesting to note that the coefficients a and c depend on $T_{CDW}$:

$$a \approx \frac{1}{T_{CDW}} \text{ and } c \approx \frac{1}{T_{CDW}^2} \qquad (4)$$

The interpretation of the anomaly of the elastic constants at a second order phase transition is based on an expansion of the free energy density in powers of the strain components developed by Rehwald [38]. The expansion is limited to the longitudinal strains $e_1$ and $e_3$ along the crystallographic a and c axes eq.1 is modified:

$$F(e_1,e_3,Q) = F_0(0,0) + \frac{1}{2}a(T-T_0)Q^2 + \frac{1}{4}cQ^4 + \frac{1}{2}C_{11,0}e_1^2 + \frac{1}{2}C_{33,0}e_3^2 + g_1e_1Q^2 + h_1e_1^2Q^2 + g_3e_3Q^2 + h_3e_3^2Q^2 \qquad (5)$$

$\frac{1}{2}C_{11,0}e_1^2 + \frac{1}{2}C_{33,0}e_3^2$ is the elastic background energy at Q=0, $g_1$ , $g_3$, $h_1$ and $h_3$ are the coupling constants. At a CDW phase transition the longitudinal strain components couples with the square power of the order parameter Q. The new elastic constants are given by:

$$C_{11} = C_{11,0} - \left(\frac{\partial^2 F}{\partial Q \partial e_1}\right)^2 \chi_Q \quad \text{and} \quad C_{33} = C_{33,0} - \left(\frac{\partial^2 F}{\partial Q \partial e_3}\right)^2 \chi_Q \qquad (6)$$

where $\chi_Q = \left(\frac{\partial^2 F}{\partial Q^2}\right)^{-1}$ denotes the order parameter susceptibility. Finally the new elastic constants are given by

$$C_{11} \approx C_{11,0} - 2\frac{g_1^2}{c} \quad \text{and} \quad C_{33} \approx C_{33,0} - 2\frac{g_3^2}{c} \qquad (7)$$

It results that a decrease of the elastic constants $C_{11}$ and $C_{33}$ occurs at a second order phase transition. An exemple of the decrease of the elastic constant $C_{33}$ observed at the CDW phase transition in $TbTe_3$ is given in Fig1(b). Furthermore from the value of c given in eq.4; it results that the decrease of the elastic constant $\Delta C_{11}$ at the transition increases as the square of the temperature $T_{CDW}$ of the phase transition

$$\Delta C_{11} \approx g_1^2 T_{CDW}^2 \quad \text{and} \quad \Delta C_{33} \approx g_3^2 T_{CDW}^2 \qquad (8)$$

In contrast the specific heat jump per unit volume at the transition increases as $T_{CDW}$ eq.2.

*Below $T_C$*

In the ordered phase, consequently to the coupling terms $h_1 e_1^2 Q^2$ and $h_3 e_3^2 Q^2$ the temperature dependence of the elastic constants $C_{11}$ and $C_{33}$ follows the temperature dependence of the square of the order parameter (black dotted curve in Fig. 1 (b) as explained by Rehwald [38]

$C_{11} \sim h_1 Q^2$ and $C_{33} \sim h_3 Q^2$ (9)

### 2.3 *Landau theory of a first order phase transition*

A cubic term is included by Mc Millan to describe the first order incommensurate-commensurate phase transition [34]. The Landau free interaction energy has the form (taking into account only the $C_{11}$ mode):

$$F(e_i, Q) = F_0 + \frac{1}{2}a(T - T_0)Q^2 - bQ^3 + \frac{1}{4}cQ^4$$
$$+\frac{1}{2}C_{11,0}e_1^2 + g_1 e_1 Q^2 \qquad (10)$$

Minimizing energy F with respect to Q one observes than three temperatures are defined [36]:

$T_0$ is the transition equilibrium temperature.

Below $T_H \sim T_0 + 9b^2/4ac$ the state is metastable.

Below $T_C \sim T_0 + 2b^2/ac$ the order parameter Q increases discontinuously in the ordered phase.

Thermal hysteresis is given by $T_H - T_C \sim b^2/ac$. Below $T_C$, the solution of $\frac{\partial F}{\partial Q} = 0$ gives the order parameter

$$Q \approx \frac{b}{c} + \sqrt{\frac{a}{c}}\left[T_C - T\right]^{0.5} \qquad (11)$$

Above $T_H$ the order parameter is zero Q=0. The small thermal hysteresis $T_H - T_C$ induces that the cubic coefficient b is small compared to the coefficient of the quartic term c.

$Q(0)^2 \approx \frac{a}{c} T_C$ is deduced from eq.11. It results that $(T_H - T_C)/T_C \sim b^2/c^2$ taking Q(0)=1.

A small hysteresis ($T_H$-$T_C$) of about 1 K with $T_C$~80 K in the case of $Lu_5Ir_4Si_{10}$ yields a ratio b/c ~0.1 [36].

The decrease of the elastic stiffness is now given by

$$C_{11} \approx C_{11,0} - 2\frac{g_1^2}{c}\frac{1}{1-(b/c)^2} \qquad (12)$$

If we neglect b/c in eq. 12 the decrease of the velocity of the $C_{11}$ is similar to the decrease obtained at a second order phase transition.

## 3 Results

*Specific heat jump per unit volume and sound velocity decrease at the CDW phase transition.*

The specific heat jumps measured and the decrease of the velocity of the longitudinal modes at the CDW (or SDW) phase transitions in several materials are reported in Tables 1 and 2. They are shown as a function of the temperature of the phase transition in Fig.2 and 3.

a) Quasi-one-dimensional conductors $K_{0.3}MoO_3$ exhibit a Peierls Transition at 180 K. The thermodynamic (specific heat and Young Modulus) properties of these materials have been studied in details [11-14]. This materials is a model for the quasi-one dimensional conductors.

b) Transition metal dichalcogenides 2H $NbSe_2$ and $TiSe_2$ :

2H-$NbSe_2$ is known to be an archetype layered transitional metal dichalcogenide superconductor with a superconducting transition temperature of 7.3 K. Quasi-two-dimensional layered transition metal dichalcogenides (TMDs) have been the subject of intense research owing to their rich electronic properties resulting from lower dimensionality. An upper limit of 0.5J/molK has been estimated on the magnitude of any anomaly in the heat specific heat associated with the 30 K transition [15, 16]. A discontinuity of velocity ΔV/V ~0.0005 in the longitudinal $C_{11}$ mode propagating in the basal plane is found at the 30 K transition [16]. A large discontinuity (ΔV/V=0.05) of the velocity of the elastic constant $C_{11}$ of $TiSe_2$ is measured at 200K [17,18].

c) Rare earth tritellurides $TbTe_3$, $ErTe_3$ and $HoTe_3$ :

The family of rare earth tritellurides $RTe_3$ (R being an element of the Lanthanide group is a model of a two dimensional charge density system [19]. The $RTe_3$ compounds crystallize in the orthorhombic structure described by the Cmcm space group. Lattice consists of stacked Te layers alternating with RTe layers with unit cell with a large anisotropy in the lattice parameters a~c ~4 Å and b~25 Å [19]. Specific heat anomalies $\Delta C_P$ ~1 and 3 J/molK and a large decrease of the velocity $\Delta V/V$~0.02 of the longitudinal $C_{33}$ mode are observed at the upper CDW phase transition in $ErTe_3$, $TbTe_3$ and $HoTe_3$ [21-23].

d) $Lu_5Ir_4Si_{10}$ and $LaAgSb_2$ :

The intermetallic compound $Lu_5Ir_4Si_{10}$ exhibits a weakly first-order (small thermal hysteresis of 1 K) charge density wave transition (CDW) at $T_{CDW}$ ~ 80 K associated with a commensurate lattice modulation along the c axis with a seven-unit cell period [25]. The first order CDW phase transition observed in $Lu_5Ir_4Si_{10}$ is characterized by a large specific heat jump, changes in the thermal expansion coefficients, and a sudden drop of the electric resistivity [25]. But the decreases of the longitudinal $C_{11}$ and $C_{33}$ elastic constants exhibit at the phase transition an apparent second order behavior [27] and the mean field specific contribution $\Delta C_P^{MF}$ ~ 6 J/molK is evaluated [25]. A charge density wave transition is observed in $LaAgSb_2$ at 210 K [28] no elastic constant measurements are reported.

e) Spin Peierls compound $CuGeO_3$ and Cr :

The quasi-one dimensional antiferromagnet CuGeO3 has been intensively studied. This compound is the first inorganic material to exhibit a spin Peierls (SP) transition which has been essentially found among the 1D magnetic organic systems. The SP transition is driven by magnetoelastic interaction between the 1D AF chains and the 3D phonon field in the lattice which opens a finite energy gap in the spin excitation spectrum by the dimerization of the lattice. The SP ground state has been demonstrated by the disappearance of the magnetic susceptibilities below $T_{SP}$=14.3 K and the observation by x-ray and neutron diffraction experiments of superstructure reflections [28]. A specific heat jump $\Delta C_P$ ~ 2 J/molK and a velocity decrease $\Delta V/V$~0.0005 are observed at $T_{SP}$=14.3 K [29]. Specific heat jump [31] and velocity decrease [32] observed in Cr at the SDW phase Temperature $T_N$=310 K are in agreement with the mean field theory and the results satisfy the relations of Ehrenfest type [33]. All the experimental data obtained with the CDW systems can be described reasonably by the red dotted line :

$$\Delta C_P / V_m = 2.8\times10^{-5} T_{CDW} \qquad (17)$$

In contrast the spin density wave systems ($CuGeO_3$ and Cr) are described by the black dotted line :

$$\Delta C_P / V_m = 1.4 \times 10^{-4} T_{CDW} \qquad (18)$$

having a coefficient five times larger. Such a lineal temperature dependence is in agreement with the Landau approach eq. 2. A remarkable feature is that ΔV/V show a $T_{CDW}^2$ dependence The steplike decreases ΔV/V of the velocity of the longitudinal $C_{11}$ or $C_{33}$ mode measured at the CDW (SDW) phase transition show a $T_{CDW}^2$ (red dotted line ) observed for all the materials :

$$\Delta V / V = 6.2 \times 10^{-7} T_{CDW}^2 \qquad (1)$$

Such a temperature dependence in agreement with the Landau approach eq. 8 is verified by quasi one dimensional conductors $K_{0.3}MoO_3$ , transition metal dichalcogenide 2H-$NbSe_2$, rare earth tritellurides $TbTe_3$, $ErTe_3$ and $HoTe_3$ and spin density wave systems $CuGeO_3$ and Cr.

## 4 Conclusion

It appears that the amplitude of the specific heat at the CDW (SDW) phase transition increases with increasing $T_{CDW}$ and a tendency to a lineal temperature dependence is found with several materials. In the same manner the amplitude of the discontinuity of the longitudinal elastic constants increases with increasing $T_{CDW}$, a quadratic dependence is verified. Nevertheless this law can not be verified with all the real materials. Thermodynamic properties of several CDW systems follow the classical mean field BCS type behavior. However departures from the mean field behavior of the thermodynamic properties attributed to fluctuations are observed with most of the transition metal dichalcogenides and many other mateials showing a CDW phase transition.

**References**

[1] G. Grüner,

**Density Waves in Solids**

ed. D. Pines (Addison-Wesley) 1994 and Rev. Mod. 60,1129 (1988).

[2] H. Frölich, Poc.R. Soc. London A223, 296, (1954).

[3] J. W. Brill, "**Elastic properties of low dimensional materials** in Chap.10 in Handbook of Elastic Properties of Solids (VolumeII) Liquid and Gases, Levy, Bass, and Stern, Eds, Academic Press, (2001).

[4] P. Monceau, **Electronic crystals**: An experimental overview, Advances in Physics, Vol 61, 325-581, (2012).

[5] M. D. Johannes, I. Mazin,

**Fermi Surface nesting and the origin of charge density waves in metals**

Phys. Rev B 77 165135 (2008).

[6] X. Zhu, J. Guo, J. Zhang, E. W. Plummer, J. D. Guo

**Classification of charge density waves based on their nature**

Proc. Natl. Acad. Sci. 112 (2015) 2367-2371

[7] X. Zhu, J. Guo, J. Zhang, E. W. Plummer,

**Misconception associated with the origin of charge density waves**

Adv. Phys. 2 622-640 (2017).

[8] S. Manzeli, D.Ovchinnikov, D. Pasquier, O. V. Yazyev, A. Kis

**2D transition metal dichacogenides**

Nature Reviews/Materials 2 (2017) 17033.

[9 J. A. Wilson, F. Di Salvo, S. Mahajan
**Charge density waves and superlattices in the metallic layered transition metal**
Advances in Physics 24   117 (1975).

[10] K. Rossnagel
**On the origin of charge density waves in selected layered transition metal dichalcogenides**
Journal of Physics: Condensed Matter 23, 213001 (2011).

[11] J. W. Brill, M. Chung, Y.-K. Kuo, E. Figueroa, G. Mozurkewich
**Thermodynamics of the Charge-Density-Wave Transition in Blue Bronze,**
Phys. Rev. Lett. 74 1182 (1995) .

[12] J. A. Aronovitz, P. Goldbart, G. Mozurkewich
**Elastic singularities at the Peierls transition**
Phy. Rev. Lett 64, 2799-2802 (1990)

[13] S. Tomic, K. Biljakovic, D. Djurek, J. R . Cooper, P. Monceau, A. Meerschaut
**Calorimetry study of the phase transition in Niobium Triselenide $NbSe_3$**
Solid. State. Commun. 38, 109-112 (1981).

[14] J. W. Brill, N. P. Ong,
**Young's modulus of $NbSe_3$**
Solid State Commun. 25, 1075-1078 (1978).

[15] V. Eremenko, V. Sirenko, V. Ibulaev, J. Bartomomé, A. Arauzo, G. Reményi,
**Heat capacity, thermal expansion and pressure derivative of critical temperature at the superconducting and charge density wave (CDW) in NbSe2**
Physica 469, 259-264 (2009).

[16] M. H. Jericho, A. M. Simpson, R. F. Frindt,
**Velocity of ultrasonic waves in 2H-$NbSe_2$, 2H-$TaS_2$ and 1T-$TaS_2$**
Phys. Rev B 22, 4907-4914 (1980).

[17] A. Caillé, Y. Lepine, M. H. Jericho, A. M. Simpson,
**Thermal expansion, ultrasonic velocity and attenuation measurements in $TiS_2$,$TiSe_2$ and $TiS_{0.5}Se_{1.5}$**
Phys. Rev. B 28, 5454 (1983).

[18] R. A. Craven, F. J. Di Salvo, F. S. L. Hsu,
**Mechanism for the 200 K transition in $TiSe_2$: A measurement of the specific heat**
Solid State Commun. 25, 39-42 (1978).

[19] N. Ru, C. L. Condron,G. Y. Margulis, K. Y. Shin, J. Laverock, S. B. Dugdale, M. F. Toney, I. R. Fisher
**Effect of chemical pressure on the charge density wav transition in rare earth tritellurides $RTe_3$**
Phys. Rev. B 77 035114 (2008).

[20] V. Brouet, W. L. Yang, X. J. Zhou, Z. Hussain, R. G. Moore, R. He, D. H. Lu, Z. X. Shen, J. Laverock, S. B. Dugdale, N. Ru, I. R. Fisher,
**Angle resolved photoemission study of the evolution, of band structure and charge density wave properties in $RTe_3$ (R=Y,La, Ce, Sm, Gd, Tb and Dy).**
Phys.Rev. B 77 235104 (2008).

[21] M. Saint-Paul, C. Guttin, P. Lejay, G. Remenyi, O. Leynaud, P. Monceau,
**Elastic anomalies at the charge density wave transition in $TbTe_3$.**
Solid State Commun.233, 24-29 (2016) .

[22] M. Saint-Paul, G. Reményi, C. Guttin, P. Lejay, P. Monceau,
**Thermodynamic and critical properties of the charge density wave system $ErTe_3$**
Physica B 504, 39-46 (2017).

[23] M. Saint-Paul, C. Guttin, P. Lejay, O. Leynaud, P. Monceau,
**Elastic anomalies at the charge density wave transition in $HoTe_3$.**
International Journal of Modern Physics B 32, 1850249 (2018).

[24] S. Van Smaalen, M. Shaz, L. Palatinus, P. Daniels, F. Galli, G. J. Nieuwenhuys, J. A. Mydosh,
**Multiple charge-density-waves in $R_5Ir_4Si_{10}$ (R=Ho, Er, Tm, and Lu)**
Phys. Rev. B 69 014103 (2004).

[25] Y.-K. Kuo, C. S. Lue, F. H. Hsu, H. H. Li, H. D. Yang,
**Thermal properties near the charge-density-wave transition**
Phys. Rev. B 64 125124 (2001) .

[26] B. Mansart, M. J. G. Cottet, T. J. Penfold, S. B. Dugdale, R. Tediosi, M. Chergui, F. Carbone,
**Evidence for a Peierls phase transition in a three-dimensional multiple charge-density waves in solid.**
Proc. Natl. Acad. Sci. (2011)

[27] M. Saint-Paul, C. Opagiste, C. Guttin,
**Elastic properties at the charge density wave transition in $Lu_5Ir_4Si_{10}$**
Journal of Physics and Chemistry of solids (to be published)

[28] S. L. Bud'ko, S. A. Law, P. C. Canfield,G. D. Samolyuk M. S. Torikachvili G. M. Schmiedeshoff,
**Thermal expansion and magnetostriction of pure and doped RAgSb2 (R=Y, Sm, La) single crystals.**
Journal of Physics: Condensed Matter, vol. 20, 115210 (2008).

[29] J. P. Pouget, L. P. Regnault, M. Ain, B. Hennion, J. P. Renard, P. Veillet, G. Dhalenne, and A. Revcolevschi,
**Structural evidence for a spin Peierls ground state in the quasi one dimensional compound $CuGeO_3$**
Phys. Rev. Lett. 72, 4037 (1994).

[30] M. Saint-Paul, G. Reményi, N. Hegmann, P. Monceau, G.Dhalenne, A. Revcolevschi,
**Ultrasonic Study o magnetoelastic effects in the spin -Peierls state of CuGeO3**
Phys. Rev. B 52, 15298 (1995).

[31]R. Weber, R. Street,

**The heat capacity anomaly of chromium at 311 K (antiferromagnetic to paramagnetic transition**

Journal of Physics F: Metal Physics 2 873-877 (1972).

[32] D. I. Bolef, J. De Klerk,

**Anomalies in the elastic constants and thermal expansion of chromium single crystals**

Phys. Rev B 129, 1063-1067 (1963).

[33] M. Saint-Paul and P. Monceau,

**Survey of the thermodynamic properties of the charge density wave systems**

Advances in condensed matter (Hindawi) (2019) 2138264 doi /10.1155/2019/2138264

[34] W. L. McMillan,

**Landau theory of charge density waves in transition-metal dichalcogenides**

Phys. Rev. B 12 1187-1196 (1975).

[35] D. Allender, J. W. Bray, J. Bardeen,

**Theory of fluctuation superconductivity from electron-phonon interactions in pseudo-one-dimensional systems**

Phys. Rev. B 9, 119 (1974).

[36 L. Landau and E. Lifshitz, Statistical Physics, Pergamon Press, 1968

and K. Binder,

**Theory of first order phase transitions**

Rep. Prog. Phys. 50 783859 (1987).

[37] L. R. Testardi,

**Elastic modulus, thermal expansion, and specific heat at a phase transition**

Phys. Rev. B 12 3849-3853 (1975).

[38] W. Rehwald,

**The study of structural phase transitions by means of ultrasonic experiments**

Adv. Phys. 22 721-755(1973).

**Table 1 Anomalies of Specific heat and sound velocity at the CDW phase transition. Molar volume $V_m$.**

| **Materials** | | **Specific heat jump** | **Velocity decrease** |
|---|---|---|---|
| **Quasi-one-dimensional conductors** | | | |
| **$K_{0.3}MoO_3$**<br>**$V_m=3.6\times10^{-5}$ m$^3$**<br><br>**$NbSe_3$**<br>**Vm=$4.1\times10^{-5}$ m$^3$** | **$T_{CDW}$=180 K**<br><br>**$T_{CDW}$=145K** | **$\Delta C_P$=3 J/molK [11]**<br><br>**$\Delta C_P$=4J/molK [13]** | **$\Delta V/V$=0.01**<br>**(Young modulus[11])**<br>**Y~250 Gpa**<br><br>**$\Delta V/V$=0.0003 [14]** |
| **Transition metal dichalcogenides**<br><br>**2H-$NbSe_2$**<br>**$V_m=4x10^{-5}$ m$^3$**<br><br>**$TiSe_2$**<br>**$V_m=3.96x10^{-5}$ m$^3$** | **$T_{CDW}$=30 K**<br><br>**$T_{CDW}$=200 K** | **$\Delta C_P$=0.5J/molK [15]**<br><br>**$\Delta C_P$=1J/molK [18]** | **$\Delta V/V$=0.0005 [16]**<br><br>**$\Delta V/V$=0.05 [17]** |
| **Rare earth tritellurides**<br>**$TbTe_3$**<br>**$V_m=7.1\times10^{-5}$ m$^3$**<br><br>**$ErTe_3$**<br>**$V_m=7.04\times10^{-5}$ m$^3$**<br><br>**$HoTe_3$** | **$T_{CDW}$=330 K**<br><br>**$T_{CDW}$=260 K**<br><br>**$T_{CDW}$=280 K** | **$\Delta C_P$=3 J/molK [21]**<br><br>**$\Delta C_P$=1 J/molK [22]** | **$\Delta V/V$=0.01($C_{33}$ mode) [21]**<br>**$\Delta V/V$=0.015($C_{33}$ mode)[22]**<br>**$\Delta V/V$=0.025 ($C_{33}$ mode)[23]**<br>**$C_{33}$~50 GPa** |
| **Intermetallic compound** | | **mean field** | |
| **$Lu_5Ir_4Si_{10}$** | | **contribution** | **$\Delta V/V=0.005$ ($C_{11}$ mode)** |
| | **$T_{CDW}$~80 K** | **$\Delta C_P^{MF}$=5J/molK [25]** | **$C_{11}$~$C_{33}$ ~230GPa[27]** |
| **$V_m=2\times10^{-4}$ m$^3$** | | | |
| **LaAgSb2** | | | |
| **$V_m=6.3x10^{-5}$ m$^3$** | **$T_{CDW}$=210 K** | | |
| | | **$\Delta C_P$=0.5 J/molK [28]** | |

**Table 2 Anomalies of Specific heat and sound velocity at the CDW (SDW) phase transition. Molar volume $V_m$.**

| | | | |
|---|---|---|---|
| **Spin Peierls compound** | | | |
| **$CuGeO_3$** | **$T_{SDW}$=14 K** | **$\Delta C_P$=2 J/molK [30]** | **$\Delta V/V=0.0005$ ($C_{33}$ mode)[30]** |
| **$V_m=3.6x10^{-5}$ m$^3$** | | | |
| **Spin density wave transition (antiferromagnetic transition)** | | | |
| **Cr** | **$T_{SDW}$=310 K** | | |
| **$V_m=7.2x10^{-6}$ m$^3$** | | | |
| | | **$\Delta C_P$=3 J/molK [31]** | **$\Delta V/V$=0.05 ($C_{33}$ mode)[32]** |

## Figure captions

**Fig. 1 color on line**

(a) Heat specific jump at the CDW phase transition in $TbTe_3$ [21].

(b) Decrease of the velocity of the longitudinal $C_{33}$ mode propagating along the c axis in $TbTe_3$ [21]. The black line is the temperature dependence of the background contribution above the transition. The dotted black line is the temperature dependence of the velocity in the ordered phase involving the temperature dependence of the order parameter. The red symbols correspond to the attenuation of the elastic wave at 15 MHz in this case [21].

**Fig.2 color on line**

Heat specific heat discontinuities $\Delta C_P/V_m$ per unit volume ( $MJ/Km^3$ ) at the CDW phase transition as a function of the temperature of the phase transition $T_{CDW}$.
Quasi one dimensional conductors $NbSe_3$ (blue circle) $T_{CDW}$=145K [13]; $KoMoO_3$ (green circle) T=180 K [11]; Transition metal dichalcogenides : 2H-$NbSe_2$ (red up triangle) $T_{CDW}$=30 K [15]; $TiSe_2$ (blue up triangle) $T_{CDW=}$200 K [18]; Rare earth tritellurides $ErTe_3$ (black down triangle) $T_{CDW}$=260 K [22]; $TbTe_3$ (pink down triangle) $T_{CDW}$=330 K [21]; $HoTe_3$ (blue down triangle) $T_{CDW}$ =280 K. Intermetallic compounds $Lu_5Ir_4Si_{10}$ (red square) $T_{CDW}$=80 K [25]; $LaAgSb_2$ (pink square) $T_{CDW}$=210 K [28] ; Spin Peierls systems $CuGeO_3$ (blue diamond) $T_{SDW}$=14 K [29] chromium Cr (light blue diamond) $T_{SDW}$=310 K [31];

**Fig. 3 color on line**

Relative velocity decrease of the velocity ($\Delta V/V$) the longitudinal $C_{11}$ and $C_{33}$ modes at the CDW phase transition as a function of the temperature of the phase transition $T_{CDW}$.
Quasi one dimensional conductors $NbSe_3$ (red up triangle) $T_{CDW}$=145K[14]; $KoMoO_3$ (green circle) T=180 K [11]; Transition metal dichalcogenides : 2H-$NbSe_2$ (blue square) $T_{CDW}$=30 K [16]; $TiSe_2$ (blue up traingle) $T_{CDW=}$200 K [17]; Rare earth tritellurides $ErTe_3$ (black down triangle) $T_{CDW}$=260 K [22], $TbTe_3$ (pink down triangle) $T_{CDW}$=330 K [21], $HoTe_3$ (blue down triangle) [23]. Intermetallic compounds $Lu_5Ir_4Si_{10}$ (red square) $T_{CDW}$=80 K [27]; Spin Peierls systems $CuGeO_3$ (blue diamond) $T_{SDW}$=14 K [29]; chromium Cr (light blue diamond) $T_{SDW}$=310 K [31].

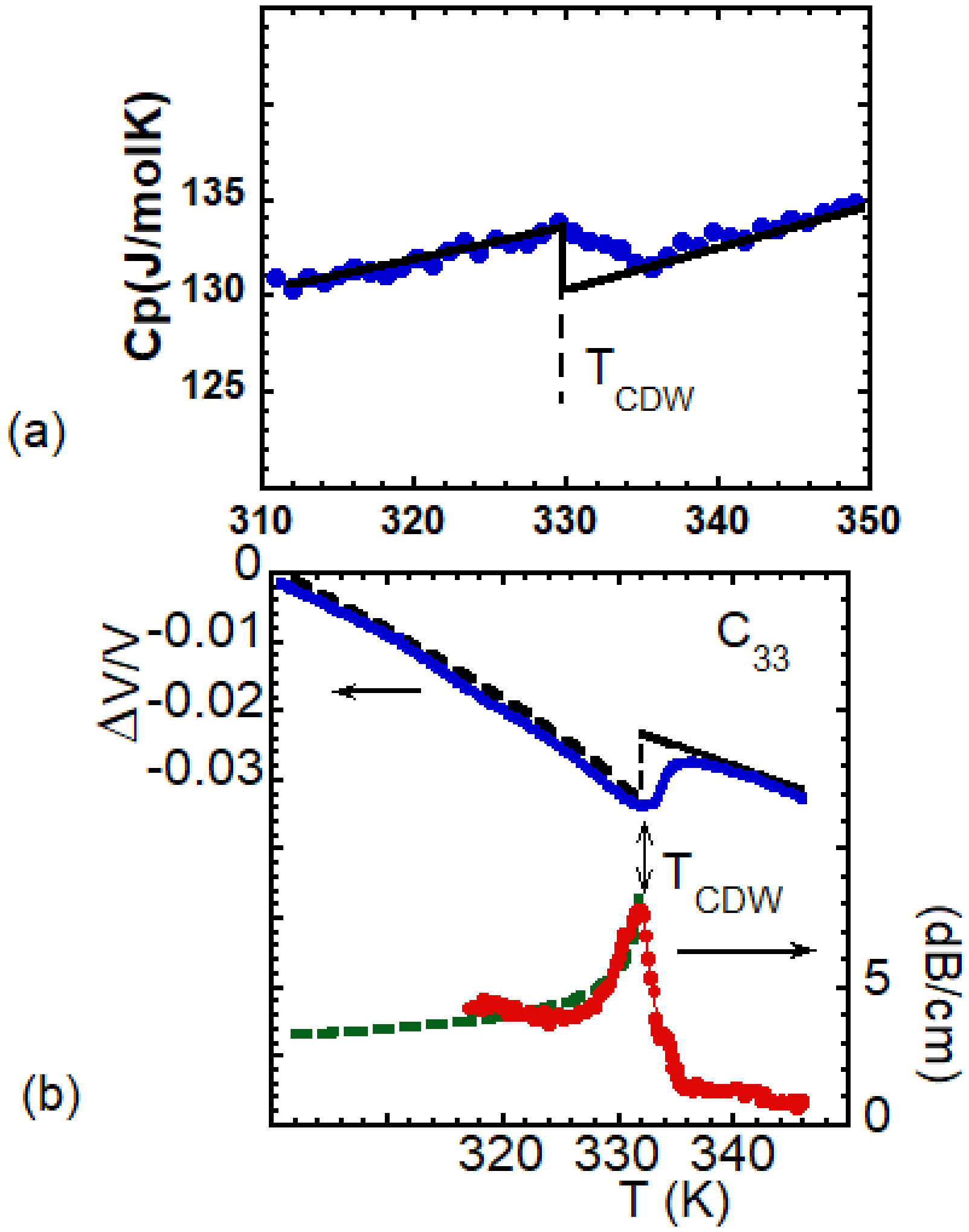

**Fig 1**

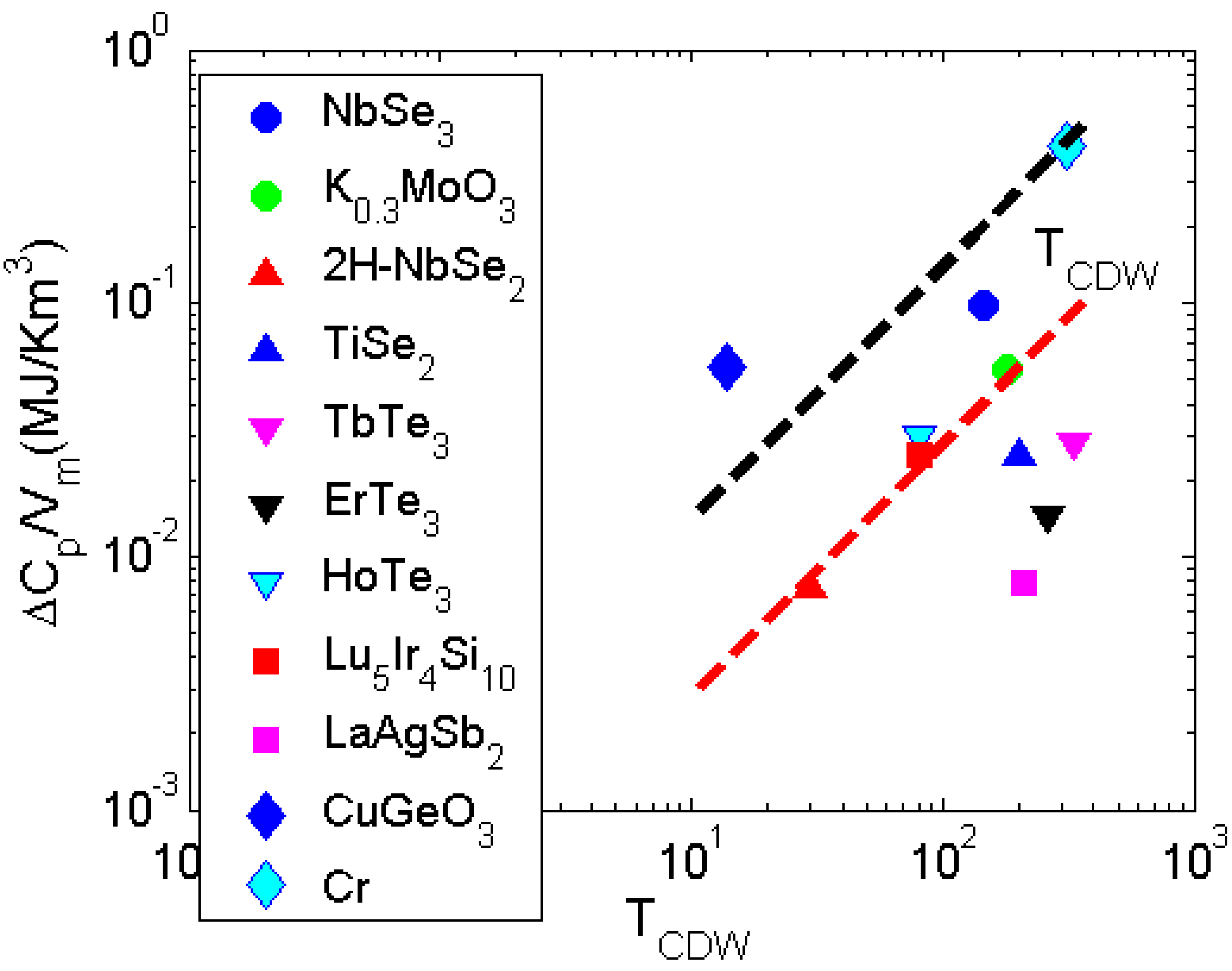

**Fig2**

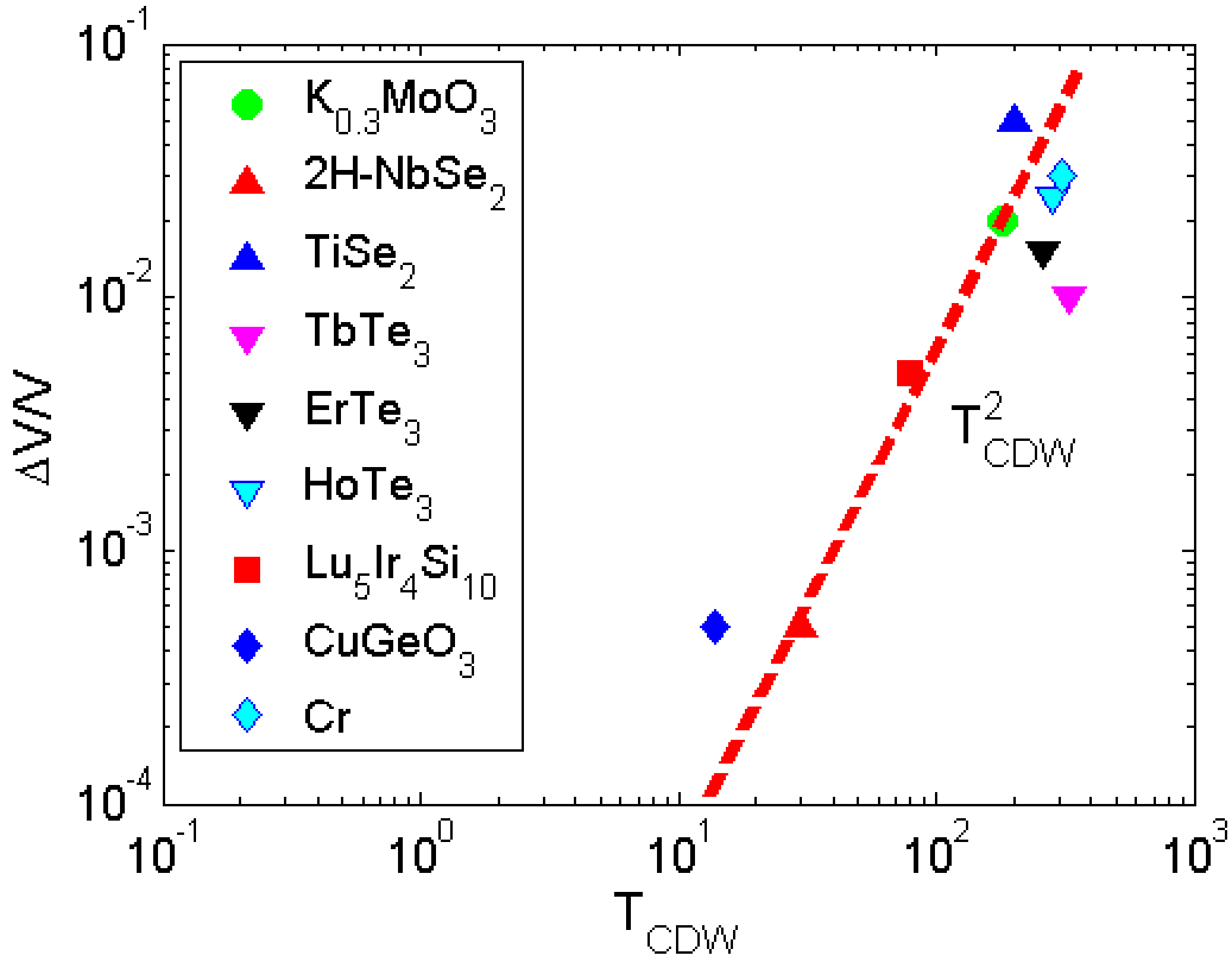